\documentclass[british,english]{revtex4-1}
\usepackage[T1]{fontenc}
\usepackage[latin9]{inputenc}
\usepackage{units}
\usepackage{mathtools}
\usepackage{amsmath}
\usepackage{amssymb}
\usepackage{stackrel}
\usepackage{esint}

\makeatletter
\usepackage[dvipsnames]{xcolor}
\usepackage[unicode=true,
 bookmarks=false,
 breaklinks=false,pdfborder={0 0 1},
 colorlinks=false]
 {hyperref}
\hypersetup{
 colorlinks,citecolor=cyan,linkcolor=blue,urlcolor=cyan}
\usepackage{breakurl}

\makeatother

\usepackage{babel}

\begin{document}
\title{Non-Equilibrium Steady State of the Lieb-Liniger model: multiple integral
representation of the time evolved many-body wave-function}
\author{Spyros Sotiriadis}
\affiliation{Department of Physics, Faculty of Mathematics and Physics, University
of Ljubljana, Ljubljana, Slovenia}
\begin{abstract}
We continue our study of the emergence of Non-Equilibrium Steady States
in quantum integrable models focusing on the expansion of a Lieb-Liniger
gas for arbitrary repulsive interaction. As a first step towards the derivation
of the asymptotics of observables in the thermodynamic and large distance
and time limit, we derive an exact multiple integral representation
of the time evolved many-body wave-function. Starting from the known but complicated 
expression for the overlaps of the initial state of a geometric quench, 
which are derived from the Slavnov formula for scalar products of Bethe states, 
we eliminate the awkward dependence on the system size and 
distinguish the Bethe states into convenient sectors. 
These steps allow us to express the rather impractical sum over Bethe states 
as a multiple rapidity integral in various alternative forms. 
Moreover, we examine the singularities of the obtained integrand 
and calculate the contribution of the multivariable kinematical poles, 
which is essential information for the derivation of the asymptotics of interest. 
\end{abstract}

\maketitle

\tableofcontents

\section{Introduction}

In an earlier work \cite{paper1}, we have studied the expansion
of a Tonks-Girardeau gas from one half to the entire confining box
and demonstrated how the emergence of Non-Equilibrium Steady States
(NESS) can be derived from the asymptotics of the many-body wave-function
in the combined thermodynamic and large distance and time limit. We
avoided using the effectively free fermionic nature of the system,
aiming to develop an exact method for the derivation of the asymptotics
in the many-body context that would be suitable for generalisation
to the genuinely interacting case of the Lieb-Liniger gas at arbitrary
repulsive interaction. A crucial intermediate step in this method is
the expression of the many-body wave-function in the form of a multiple
rapidity integral, in which the integrand is well-defined in the thermodynamic
limit. 

The method is analogous to a general method for the derivation of
asymptotics of equilibrium correlation functions in Bethe Ansatz solvable models \cite{Kitanine2005:contour,Goehmann2004:contour,Kitanine2005a:contour,Kitanine2011,Kozlowski2011,Kozlowski2014,Kozlowski2019,Goehmann2020}
(\cite{Slavnov:lecture-notes} for a recent review). The main challenge
in this type of problems is that, even though all of the ingredients
that are necessary for a formal expression of the observable of interest
are known, at least in some implicit form (Bethe states and Bethe roots,
matrix elements of the observables in the Bethe state basis), expanding
in a thermodynamically large eigenstate basis and extracting the asymptotics
of the resulting sum is very difficult. The general method to tackle
this problem consists of two main steps. The first step is to express
the observable of interest in an exact multiple integral representation \cite{Jimbo_1992,Jimbo_1995,Jimbo_1996,Kitanine2005a:contour,Kitanine2005:contour,Goehmann2004:contour,Goehmann_2005}.
The second step is to derive the asymptotics from the multiple integral
representation \cite{Kitanine_2002,Kitanine2005a:contour,Kitanine_2009b,Kitanine_2009,Kitanine_2011,Kozlowski2011,Kozlowski_2011,Kozlowski2014,Kozlowski2019}.
This method has been successfully applied to the study of asymptotics
of correlations in XXZ and Lieb-Liniger model at thermal equilibrium.

The first step is crucial as the expansion of the observable in the
Bethe eigenstate basis is not suitable for asympotic analysis. On
the contrary, the multiple integral representation is especially suitable,
since by identifying the locations of poles of the integrand in the
complex rapidity plane and performing a convenient deformation of
the integration contours, it is possible to reduce the derivation
of the asymptotics to the evaluation of multivariable pole residues.
The integrand of such multiple integral expressions generally has
the form of a determinant of a thermodynamically large matrix or products
and ratios of such determinants. 

In the present work, we are interested in the asymptotics of observables
in an out-of-equilibrium problem, more specifically, a geometric quench.
The above outlined method has not been applied to out-of-equilibrium
problems so far. Here, our goal is to generalise the method in the
form shown in \cite{paper1} to the genuinely interacting case of
the Lieb-Liniger model at arbitrary interaction $c>0$. In particular,
we focus on the time evolved many-body wave-function after the quench
and derive a multiple integral representation for it, thus accomplishing
the first step of the above general approach. Deriving the asymptotics
of this quantity in the combined thermodynamic and large distance
and time limit would allow us to establish the emergence of the NESS
and extract its dependence on the ray ratio $x/t=\xi$, providing
a way to verify the predictions of the recently introduced Generalised
HydroDynamics theory \cite{GHD_Doyon,GHD_Fagotti-Bertini}. It should
be noted that perfoming this first step for the many-body wave-function
is generally simpler than for correlation functions, because the analysis
of the latter comes with additional complications related to the complex
form of the matrix elements of field operators. On the other hand,
in the application of this method to a quantum quench problem, a different
type of complications emerge due to the fact that the initial state
overlaps, when known exactly, typically have a very complex functional
form: they are not necessarily smooth functions of the rapidity variables
but may instead be non-vanishing only for special classes of states
\cite{De_Nardis_2014,Brockmann_2014,Brockmann_2014b}. 

In the case of a geometric quench, in particular,
even though the initial state overlaps are exactly known \cite{Mossel-Caux} 
being derivable from the Slavnov formula for scalar products of Bethe states, 
their complicated form does not allow for a direct expression of the many-body wave-function 
in the form of a multiple rapidity integral. By eliminating the system size dependence
using the Bethe equations, we observe that the overlaps exhibit an 
alternating-sign behaviour. Because of this feature, a suitable splitting of the Bethe states 
into sectors is required to arrive at the desired multiple integral representation.
At the same time, the analyticity properties of the resulting integrands 
may be severely restricted. Evidently, this endangers the possibility
of performing the contour deformations that may be necessary for bringing
the integral in a form convenient for the evaluation of its asymptotics.
It is therefore crucial to examine and identify the singularities
of the integrand of the obtained multiple integral formula and, if
possible, manipulate it so as to bring it into a convenient form.
For these reasons, we express the above formula in
alternative forms and study the analytical properties of the integrand,
observing the presence of branch cuts and calculating the residue
of its multivariable kinematical poles. Based on the calculation presented in \cite{paper1}
and the above discussion, this information is expected to be important 
for the derivation of the asymptotics we are interested in.

The paper is organised as follows. We first introduce the necessary
preliminaries to our calculation: the definition of the Lieb-Liniger
model, its solution by Bethe Ansatz, the Slavnov and Gaudin formulas
for scalar products and norms of Bethe states, respectively, as well
as the definition of the quench protocol and statement of our objectives
(sec. \ref{sec:Preliminaries}). Short proof outlines and alternative
forms of some useful formulas are included for use in later discussions.
We then present in detail the derivation of a multiple integral representation
for the time evolved wave-function of the system after the quench
(sec. \ref{sec:main}), starting with a manipulation of the initial
state overlaps (subsec. \ref{subsec:Initial-state-overlaps}), using
a suitable variant of a standard complex analysis trick for passing
from the sum over Bethe states to multiple rapidity integrals and
further manipulating the resulting formula to obtain an alternative
version (subsec. \ref{subsec:Cauchy}). We derive a simple formula
for the kinematical pole residue of the integrand, a result important
for subsequent steps (sec. \ref{sec:pole}), and discuss the role of branch cut
singularities of the integrand. Lastly, we underline aspects of the
derivation that are likely to be generally valid and their relevance
for a mathematically rigorous solution of the problem of quantum dynamics
in integrable models (sec. \ref{sec:Discussion}). A heuristic derivation
of a crucial intermediate formula is presented in the appendix (app.
\ref{app:F}).

\section{Preliminaries\label{sec:Preliminaries}}

\subsection{Bethe Ansatz solution for the Lieb-Liniger model\label{subsec:Bethe_Ansatz}}

The Lieb-Liniger model for a system of $N$ interacting bosons in
a box of length $L$ with periodic boundary conditions $\Psi(+L/2)=\Psi(-L/2)$ 
is described by the Hamiltonian
\[
H=\int_{-L/2}^{+L/2}\mathrm{d}x\,\left[-\Psi^{\dagger}(x)\partial_{x}^{2}\Psi(x)+c\,\Psi^{\dagger}(x)\Psi^{\dagger}(x)\Psi(x)\Psi(x)\right]
\]
where $c$ is the interaction strength and the particle mass has been
set to $m=1/2$. In this work we will consider only the repulsive
case $c>0$. The particle number and momentum operators are respectively
\begin{align*}
N & =\int_{-L/2}^{+L/2}\mathrm{d}x\,\Psi^{\dagger}(x)\Psi(x)\\
P & =-\mathrm{i}\int_{-L/2}^{+L/2}\mathrm{d}x\,\Psi^{\dagger}(x)\partial_{x}\Psi(x)
\end{align*}

The Lieb-Liniger model is integrable i.e. its eigenstates are given
by the Bethe Ansatz. The coordinate space wave-functions of the Bethe
eigenstates are 
\begin{align}
\langle\boldsymbol{x}|\Phi(\boldsymbol{\lambda})\rangle & \coloneqq\langle x_{1},x_{2},\dots,x_{N}|\Phi(\lambda_{1},\lambda_{2},\dots,\lambda_{N})\rangle\nonumber \\
 & \propto \frac{1}{\sqrt{N!\prod_{j>i}\left[\left(\lambda_{j}-\lambda_{i}\right)^{2}+c^{2}\right]}}\sum_{\text{all perm. }\pi}(-1)^{\sigma(\pi)}\exp\left(\mathrm{i}\sum_{i=1}^{N}\lambda_{\pi_{i}}x_{i}\right)\prod_{j>i}\left(\lambda_{\pi_{j}}-\lambda_{\pi_{i}}-\mathrm{i}c\,\text{sign}(x_{j}-x_{i})\right)\label{eq:coord_BA_wf}
\end{align}
where in order to satisfy the boundary conditions, the rapidities
$\boldsymbol{\lambda}\coloneqq\{\lambda_{j}\}_{j=1}^{N}$ must satisfy
the Bethe equations (BA)
\begin{equation}
\exp\left(\mathrm{i}\lambda_{j}L\right)=\prod_{i(\neq j)}^{N}\frac{\lambda_{j}-\lambda_{i}+\mathrm{i}c}{\lambda_{j}-\lambda_{i}-\mathrm{i}c}\qquad\text{for all }j=1,2,\dots,N\label{eq:BAeqs}
\end{equation}
The latter can also be written in the following equivalent form
\begin{align}
\exp\left(\mathrm{i}Q_{i}(\boldsymbol{\lambda})\right) & =1\label{eq:BAeqs1}
\end{align}
where we defined 

\begin{align}
Q_{i}(\boldsymbol{\lambda}) & \coloneqq\lambda_{i}L+\sum_{j(\neq i)}^{N}\theta(\lambda_{i}-\lambda_{j})\label{eq:Qdef}
\end{align}
with
\begin{align}
\theta(\lambda) & \coloneqq-\mathrm{i}\log\left(\frac{\lambda-\mathrm{i}c}{\lambda+\mathrm{i}c}\right)=-\mathrm{i}\log S(\lambda)\label{eq:theta}
\end{align}
and
\begin{equation}
S(\lambda)\coloneqq\frac{\lambda-\mathrm{i}c}{\lambda+\mathrm{i}c}\label{eq:S-matrix}
\end{equation}
the two-particle S-matrix of the Lieb-Liniger model. In the repulsive
case considered here all solutions of the Bethe equations correspond
to real rapidities. The energy and momentum eigenvalues corresponding
to an eigenstate with rapidities $\boldsymbol{\lambda}$ are respectively
\begin{align}
E(\boldsymbol{\lambda}) & =\sum_{i=1}^{N}e(\lambda_{i})=\sum_{i=1}^{N}\lambda_{i}^{2}\label{eq:e}\\
P(\boldsymbol{\lambda}) & =\sum_{i=1}^{N}p(\lambda_{i})=\sum_{i=1}^{N}\lambda_{i}\label{eq:p}
\end{align}

In the hard-core boson limit $c\to\infty$ the Bethe equations simplify
to 
\[
\exp\left(\mathrm{i}\lambda_{j}L\right)=(-1)^{N-1}\qquad\text{for all }j=1,2,\dots,N
\]
with solutions
\[
\lambda_{j}=\begin{cases}
\frac{2\pi}{L}n_{j} & \text{ for }N\text{ odd}\\
\frac{2\pi}{L}\left(n_{j}+\frac{1}{2}\right) & \text{ for }N\text{ even}
\end{cases},\quad n_{j}\in\mathbb{Z}
\]
and the eigenfunctions are
\[
\langle\boldsymbol{x}|\Phi(\boldsymbol{\lambda})\rangle=\frac{1}{\sqrt{N!}}\det_{i,j}\left[\exp\left(\mathrm{i}\lambda_{j}x_{i}\right)\right]\prod_{j>i}\text{sign}(x_{j}-x_{i})
\]
This is the Tonks-Girardeau limit considered in \cite{paper1},
which will be used as a consistency check.

\subsection{Bethe state scalar products: Slavnov formula\label{subsec:Slavnov_formula}}

Overlaps between Bethe states are given by the \emph{Slavnov formula}
\cite{Slavnov}. The latter gives the scalar product $S_{N}(\boldsymbol{\mu};\boldsymbol{\lambda})\coloneqq\langle\Phi(\boldsymbol{\mu})|\Phi(\boldsymbol{\lambda})\rangle$
between two Bethe states, one of which has rapidities $\boldsymbol{\mu}$
satisfying Bethe equations while those of the other one, $\boldsymbol{\lambda}$,
are left unconstrained. The Slavnov formula is\foreignlanguage{english}{
\begin{equation}
S_{N}(\boldsymbol{\mu};\boldsymbol{\lambda})=G_{N}(\boldsymbol{\mu},\boldsymbol{\lambda})\det\left[M_{N}(\boldsymbol{\mu};\boldsymbol{\lambda})\right]_{lk}\label{eq:Slavnov}
\end{equation}
where} \foreignlanguage{english}{
\begin{equation}
G_{N}(\boldsymbol{\mu},\boldsymbol{\lambda})\coloneqq\prod_{j>k}^{N}g(\lambda_{j},\lambda_{k})g(\mu_{k},\mu_{j})\prod_{j,k}^{N}h(\mu_{j},\lambda_{k})\label{eq:Slavnov1}
\end{equation}
\begin{equation}
\left[M_{N}(\boldsymbol{\mu};\boldsymbol{\lambda})\right]_{lk}\coloneqq\frac{g(\mu_{k},\lambda_{l})}{h(\mu_{k},\lambda_{l})}-r(\lambda_{l})\frac{g(\lambda_{l},\mu_{k})}{h(\lambda_{l},\mu_{k})}\prod_{m=1}^{N}\frac{f(\lambda_{l},\mu_{m})}{f(\mu_{m},\lambda_{l})}\label{eq:Slavnov2}
\end{equation}
and} \foreignlanguage{english}{the functions $g(\lambda,\lambda'),f(\lambda,\lambda'),h(\lambda,\lambda')$}
and $r(\lambda)$ are defined as\foreignlanguage{english}{
\begin{align}
g(\lambda,\lambda') & \coloneqq\frac{\mathrm{i}c}{\lambda-\lambda'}\nonumber \\
f(\lambda,\lambda') & \coloneqq\frac{\lambda-\lambda'+\mathrm{i}c}{\lambda-\lambda'}=g(\lambda,\lambda')\left(1+\frac{1}{g(\lambda,\lambda')}\right)\label{eq:gfh}\\
h(\lambda,\lambda') & \coloneqq\frac{f(\lambda,\lambda')}{g(\lambda,\lambda')}=\frac{\lambda-\lambda'+\mathrm{i}c}{\mathrm{i}c}\nonumber 
\end{align}
and 
\begin{equation}
r(\lambda)\coloneqq\mathrm{e}^{-\mathrm{i}\lambda L}\label{eq:r}
\end{equation}
The Bethe equations (\ref{eq:BAeqs}) can be expressed in terms of
the above functions as
\begin{align}
r(\lambda_{j})\prod_{i(\neq j)}^{N}\frac{f(\lambda_{j},\lambda_{i})}{f(\lambda_{i},\lambda_{j})} & =1\qquad\text{for all }j=1,2,\dots,N\label{eq:BAeqs2}
\end{align}
}

\selectlanguage{english}%
Note that the overlaps as given by the Slavnov formula are symmetric
under permutations of either set of rapidities \foreignlanguage{british}{$\boldsymbol{\lambda}$
and $\boldsymbol{\mu}$}, as they should be. To see this we first
notice that if two rapidities are exchanged then $G_{N}(\boldsymbol{\mu},\boldsymbol{\lambda})$
changes sign due to the fact that $g(\lambda',\lambda)=-g(\lambda,\lambda')$.
If we do an arbitrary permutation\foreignlanguage{british}{,} $G_{N}(\boldsymbol{\mu},\boldsymbol{\lambda})$
will pick up a sign equal to the signature of the permutation. On
the other hand, a rapidity permutation is equivalent to a permutation
of the rows or columns of the matrix $\left[M_{N}(\boldsymbol{\mu};\boldsymbol{\lambda})\right]_{lk}$
and so it also results in the same sign change for the determinant
$\det\left[M_{N}(\boldsymbol{\mu};\boldsymbol{\lambda})\right]_{lk}$,
which therefore cancels out with the sign change of $G_{N}(\boldsymbol{\mu},\boldsymbol{\lambda})$.

The Slavnov formula (\ref{eq:Slavnov}) can be written in the following
equivalent form

\begin{align}
S_{N}(\boldsymbol{\mu};\boldsymbol{\lambda}) & =\frac{1}{\det\left[g(\mu_{k},\lambda_{l})\right]_{lk}}\det\left[g^{2}(\mu_{k},\lambda_{l})\left(\prod_{m(\neq k)}^{N}f(\mu_{m},\lambda_{l})-r(\lambda_{l})\prod_{m(\neq k)}^{N}f(\lambda_{l},\mu_{m})\right)\right]_{lk}\label{eq:Slavnov-det}
\end{align}
To see this we first write $G_{N}$ as
\begin{align}
G_{N}(\boldsymbol{\mu},\boldsymbol{\lambda}) & =\frac{\prod_{j>k}^{N}g(\lambda_{j},\lambda_{k})g(\mu_{k},\mu_{j})}{\prod_{j,k}^{N}g(\mu_{j},\lambda_{k})}\prod_{j,k}^{N}f(\mu_{j},\lambda_{k})\nonumber \\
 & =\frac{1}{\det\left[g(\mu_{j},\lambda_{k})\right]_{jk}}\prod_{j,k}^{N}f(\mu_{j},\lambda_{k})\label{eq:G2}
\end{align}
where we used the \textit{Cauchy determinant} formula
\begin{align*}
\det\left[\frac{1}{\mu_{j}-\lambda_{k}}\right]_{jk} & =\frac{\prod_{j>k}^{N}(\lambda_{j}-\lambda_{k})(\mu_{k}-\mu_{j})}{\prod_{j,k}^{N}(\mu_{j}-\lambda_{k})}
\end{align*}
Next we absorb the product $\prod_{j,k}^{N}f(\mu_{j},\lambda_{k})$
in $\det\left[M_{N}(\boldsymbol{\mu};\boldsymbol{\lambda})\right]_{lk}$
using the formula
\begin{align*}
\det\left[B_{i}A_{ij}\right]_{ij} & =\left(\prod_{i=1}^{N}B_{i}\right)\det A_{ij}
\end{align*}
so that
\begin{align*}
S_{N}(\boldsymbol{\mu};\boldsymbol{\lambda}) & =\frac{1}{\det\left[g(\mu_{l},\lambda_{k})\right]_{lk}}\det\left[\prod_{j=1}^{N}f(\mu_{j},\lambda_{l})\left(\frac{g(\mu_{k},\lambda_{l})}{h(\mu_{k},\lambda_{l})}-r(\lambda_{l})\frac{g(\lambda_{l},\mu_{k})}{h(\lambda_{l},\mu_{k})}\prod_{m=1}^{N}\frac{f(\lambda_{l},\mu_{m})}{f(\mu_{m},\lambda_{l})}\right)\right]_{lk}\\
 & =\frac{1}{\det\left[g(\mu_{l},\lambda_{k})\right]_{lk}}\det\left[g^{2}(\mu_{k},\lambda_{l})\left(\prod_{m(\neq k)}^{N}f(\mu_{m},\lambda_{l})-r(\lambda_{l})\prod_{m(\neq k)}^{N}f(\lambda_{l},\mu_{m})\right)\right]_{lk}
\end{align*}
which is precisely (\ref{eq:Slavnov-det}).
\selectlanguage{british}%

\subsection{Bethe state norm: Gaudin-Korepin formula\label{subsec:Gaudin-formula}}

The Bethe Ansatz eigenstates in the standard form (\ref{eq:coord_BA_wf})
with rapidities satisfying the Bethe equations are not normalised.
Their norm is given by the \emph{Gaudin-Korepin formula} \cite{Gaudin,Korepin}
\begin{align}
\mathcal{N}(\boldsymbol{\lambda}) & \coloneqq\langle\Phi(\boldsymbol{\lambda})|\Phi(\boldsymbol{\lambda})\rangle=c^{N}\left(\prod_{m\neq\ell}f(\lambda_{m},\lambda_{\ell})\right)\,\varrho_{N,L}(\boldsymbol{\lambda})\label{eq:norm}
\end{align}
where 
\begin{align}
\varrho_{N,L}(\boldsymbol{\lambda}) & \coloneqq\det\left(\frac{\partial Q_{i}(\boldsymbol{\lambda})}{\partial\lambda_{j}}\right)\label{eq:DoS}
\end{align}
which in the thermodynamic limit expresses the density of states corresponding
to a given distribution of rapidities. The above formula for the norm
of Bethe states was conjectured by Gaudin \cite{Gaudin} and later
proved by Korepin \cite{Korepin}. Notice that it can be derived
from the Slavnov formula (\ref{eq:Slavnov}) taking the limit $\boldsymbol{\lambda}\to\boldsymbol{\mu}$
\begin{align*}
\lim_{\epsilon\to0}S_{N}(\boldsymbol{\mu},\boldsymbol{\mu}+\epsilon) & =\mathcal{N}(\boldsymbol{\mu})
\end{align*}
To see this we note that \foreignlanguage{english}{in the limit }$\boldsymbol{\lambda}\to\boldsymbol{\mu}$\foreignlanguage{english}{
the factors $r(\lambda_{l})\prod_{m=1}^{N}f(\lambda_{l},\mu_{m})/f(\mu_{m},\lambda_{l})$
}in (\ref{eq:Slavnov2})\foreignlanguage{english}{ tend to $-1$,}
since the $\boldsymbol{\lambda}$ that were previously unconstrained
now satisfy the Bethe equations too, like the $\boldsymbol{\mu}$.
As a result the poles of \foreignlanguage{english}{the functions $g(\mu_{k},\lambda_{l})=-g(\lambda_{l},\mu_{k})$
in $M_{N}(\boldsymbol{\mu};\boldsymbol{\lambda})$ at equal rapidities
are cancelled by the zeros of $1/h(\mu_{k},\lambda_{l})-1/h(\lambda_{l},\mu_{k})$
at the same points. Using in addition that
\begin{align}
G_{N}(\boldsymbol{\mu},\boldsymbol{\mu}) & =\prod_{j>k}^{N}g(\mu_{j},\mu_{k})g(\mu_{k},\mu_{j})\prod_{j,k}^{N}h(\mu_{j},\mu_{k})\nonumber \\
 & =\frac{\prod_{j>k}^{N}g(\mu_{j},\mu_{k})g(\mu_{k},\mu_{j})\prod_{j\neq k}^{N}f(\mu_{j},\mu_{k})}{\prod_{j\neq k}^{N}g(\mu_{j},\mu_{k})}\nonumber \\
 & =\prod_{j\neq k}^{N}f(\mu_{j},\mu_{k})\label{eq:Gdiag}
\end{align}
the calculation of the limit is straightforward.}

\subsection{Quench protocol and objectives}

The geometric quench problem we shall study can be described as follows.
We consider a Lieb-Liniger gas of $N$ particles, initially confined
in a box of length $L/2$; let us say in the right half of a box of
length $L$. We assume that the system lies in the ground state or
any other eigenstate $|\Phi_{0}(\boldsymbol{\mu})\rangle$ of the
Lieb-Liniger Hamiltonian restricted in this interval 
\begin{align}
H_{0} & =\int_{0}^{L/2}\mathrm{d}x\,\left[-\Psi^{\dagger}(x)\partial_{x}^{2}\Psi(x)+c\,\Psi^{\dagger}(x)\Psi^{\dagger}(x)\Psi(x)\Psi(x)\right]
\end{align}
with periodic boundary conditions $\Psi(L/2) =\Psi(0)$. 
The rapidities $\boldsymbol{\mu}$ characterising the initial state
$|\Phi_{0}(\boldsymbol{\mu})\rangle$ which is an eigenstate of $H_{0}$,
satisfy Bethe equations corresponding to system length equal to $L/2$
(BA$_{0}$)
\begin{equation}
\exp\left(\mathrm{i}\mu_{j}L/2\right)=\prod_{i(\neq j)}^{N}\frac{\mu_{j}-\mu_{i}+\mathrm{i}c}{\mu_{j}-\mu_{i}-\mathrm{i}c}\qquad\text{for all }j=1,2,\dots,N\label{eq:BAeqs0}
\end{equation}

We then time evolve the initial state under the Lieb-Liniger Hamiltonian
$H$ defined in the entire interval of length $L$ again with periodic
boundary conditions $\Psi(L)=\Psi(0)$. Generally we are interested in the asymptotics
of local observables $\hat{\mathcal{O}}(x,t)$ in the thermodynamic
limit where both $N$ and $L$ tend to infinity with fixed non-zero
density $N/L=n$, followed by the limit of large time $t$ and distance
$x$ keeping the ratio $x/t$ fixed. As discussed in more detail in
\cite{paper1}, our strategy is to focus on the quantity 
\begin{equation}
\mathcal{K}(\boldsymbol{\mu};\boldsymbol{z};x,t)\coloneqq\langle\boldsymbol{z}|\mathrm{e}^{+\mathrm{i}Px-\mathrm{i}Ht}|\Phi_{0}(\boldsymbol{\mu})\rangle\label{eq:K}
\end{equation}
i.e. the time evolved may-body wave-function projected onto a local
basis $\langle\boldsymbol{z}|$, and to evaluate it in the above combined
limit. Our objective, in the present work, is to derive an exact multiple integral
formula for $\mathcal{K}(\boldsymbol{\mu};\boldsymbol{z};x,t)$ as
a first step of this program. Similar multiple integral representations of
many-body wave-functions in integrable models under special initial
conditions have been also studied in \cite{Tracy_2008,Feher2019},
even though those works focus on the many-body propagator or Green's
function which formally solves the time-dependent problem when the initial
positions of the particles are given, while here we consider instead
the case where the initial state is an eigenstate of the pre-quench
Hamiltonian $H_0$.

\section{Derivation of multiple integral representation for the time evolved
many-body wave-function\label{sec:main}}

We start by formally expanding the initial state in the post-quench
basis, introducing a resolution of the identity in terms of Bethe
eigenstates 
\begin{align}
\mathrm{e}^{+\mathrm{i}Px-\mathrm{i}Ht}|\Phi_{0}(\boldsymbol{\mu})\rangle & =\sum_{\boldsymbol{\lambda}:\text{ BA}}\mathrm{e}^{+\mathrm{i}P(\boldsymbol{\lambda})x-\mathrm{i}E(\boldsymbol{\lambda})t}\frac{|\Phi(\boldsymbol{\lambda})\rangle\langle\Phi(\boldsymbol{\lambda})|}{\langle\Phi(\boldsymbol{\lambda})|\Phi(\boldsymbol{\lambda})\rangle}|\Phi_{0}(\boldsymbol{\mu})\rangle\nonumber \\
 & =\sum_{\boldsymbol{\lambda}:\text{ BA}}\mathrm{e}^{+\mathrm{i}P(\boldsymbol{\lambda})x-\mathrm{i}E(\boldsymbol{\lambda})t}\frac{\mathcal{M}(\boldsymbol{\lambda};\boldsymbol{\mu})}{\mathcal{N}(\boldsymbol{\lambda})}|\Phi(\boldsymbol{\lambda})\rangle\label{eq:int1}
\end{align}
where the sum runs over all solutions of the Bethe equations for $N$
particles in length $L$, the eigenstate overlaps and norms are
\begin{align*}
\mathcal{M}(\boldsymbol{\lambda};\boldsymbol{\mu}) & \coloneqq\langle\Phi(\boldsymbol{\lambda})|\Phi_{0}(\boldsymbol{\mu})\rangle\\
\mathcal{N}(\boldsymbol{\lambda}) & \coloneqq\langle\Phi(\boldsymbol{\lambda})|\Phi(\boldsymbol{\lambda})\rangle
\end{align*}
and the energy and momentum eigenvalues are given by (\ref{eq:e})
and (\ref{eq:p}).

\subsection{Initial state overlaps\label{subsec:Initial-state-overlaps}}

The first step of our study is the calculation of the overlaps $\mathcal{M}(\boldsymbol{\lambda};\boldsymbol{\mu})$.
Even though for a general quantum quench this is typically a hard
task, luckily the initial state overlaps for the geometric quench
are exactly known and given by the Slavnov formula with different
Bethe equations imposed on the two Bethe states, as was observed in
\cite{Mossel2010,Mossel-Caux}.

Let us first write the overlaps in terms of the coordinate space wave-functions
of the Bethe Ansatz eigenstates 
\[
\mathcal{M}(\boldsymbol{\lambda};\boldsymbol{\mu})=\int_{0}^{L/2}\mathrm{d}\boldsymbol{x}\;\langle\Phi(\boldsymbol{\lambda})|\boldsymbol{x}\rangle\langle\boldsymbol{x}|\Phi_{0}(\boldsymbol{\mu})\rangle
\]
Next we observe that the functional form (\ref{eq:coord_BA_wf}) of
the Bethe wave-functions is independent of the system size: the Bethe
Ansatz eigenstates depend on the latter only implicitly through the
rapidities and because these are solutions of the Bethe equations
(\ref{eq:BAeqs}) which are the ones that depend explicitly on the
system size. Therefore both pre- and post-quench eigenstates are given
by the same functions of the rapidity variables, the difference being
that the rapidities $\boldsymbol{\mu}$ and $\boldsymbol{\lambda}$
take different values as they satisfy different equations, more precisely,
the Bethe equations corresponding to system sizes $L/2$ and $L$,
respectively. 

This means that in order to calculate $\mathcal{M}(\boldsymbol{\lambda};\boldsymbol{\mu})$
we can use the Slavnov formula (\ref{eq:Slavnov}) that gives the
scalar product $S_{N}$ between two Bethe states, one with rapidities
satisfying Bethe equations and the other with unconstrained rapidities.
For our purposes we set the state described by the fixed rapidities
to be an eigenstate of the half system, therefore we impose the corresponding
Bethe equations (BA$_{0}$) for the rapidities $\boldsymbol{\mu}$
\begin{equation}
\mathrm{e}^{-\mathrm{i}\mu_{i}L/2}\prod_{j(\neq i)}^{N}\left(\frac{\mu_{i}-\mu_{j}+\mathrm{i}c}{\mu_{i}-\mu_{j}-\mathrm{i}c}\right)=\mathrm{e}^{-\mathrm{i}\mu_{i}L/2}\prod_{j(\neq i)}^{N}\frac{f(\mu_{i},\mu_{j})}{f(\mu_{j},\mu_{i})}=\exp\left(-\mathrm{i}Q_{i}^{(0)}(\boldsymbol{\mu})\right)=1\label{eq:BAeqs_L/2}
\end{equation}
where we define $Q_{i}^{(0)}(\boldsymbol{\lambda})$ as in (\ref{eq:Qdef})
but for the half system i.e. replacing $L$ by $L/2$ 
\begin{equation}
Q_{i}^{(0)}(\boldsymbol{\lambda})\coloneqq\lambda_{i}L/2+\sum_{j(\neq i)}^{N}\theta(\lambda_{i}-\lambda_{j})\label{eq:Q0def}
\end{equation}
The rapidities $\boldsymbol{\lambda}$ of the other state, which are
generally unconstrained, are set to satisfy the Bethe equations for
the entire system of length $L$ (BA)

\begin{equation}
\mathrm{e}^{-\mathrm{i}\lambda_{i}L}\prod_{j(\neq i)}^{N}\left(\frac{\lambda_{i}-\lambda_{j}+\mathrm{i}c}{\lambda_{i}-\lambda_{j}-\mathrm{i}c}\right)=\mathrm{e}^{-\mathrm{i}\lambda_{i}L}\prod_{j(\neq i)}^{N}\frac{f(\lambda_{i},\lambda_{j})}{f(\lambda_{j},\lambda_{i})}=\exp\left(-\mathrm{i}Q_{i}(\boldsymbol{\lambda})\right)=1\label{eq:BAeqs_L}
\end{equation}
In particular, note that choosing the fixed rapidities $\boldsymbol{\mu}$
to satisfy (\ref{eq:BAeqs_L/2}) means that the function \foreignlanguage{english}{$r$
in }(\ref{eq:Slavnov})\foreignlanguage{english}{ is equal to $r(\mu)=\mathrm{e}^{-\mathrm{i}\mu L/2}$,
because the system size entering the definition (\ref{eq:r}) is now
$L/2$. }

Noting that the fixed rapidities $\boldsymbol{\mu}$ enter in (\ref{eq:Slavnov})
in the bra-state and $\boldsymbol{\lambda}$ in the ket-state, we
have \foreignlanguage{english}{
\begin{equation}
\mathcal{M}(\boldsymbol{\lambda};\boldsymbol{\mu})^{*}=S_{N}(\boldsymbol{\mu};\boldsymbol{\lambda})=\prod_{j>k}^{N}g(\lambda_{j},\lambda_{k})g(\mu_{k},\mu_{j})\prod_{j,k}^{N}h(\mu_{j},\lambda_{k})\det\left[\frac{g(\mu_{k},\lambda_{l})}{h(\mu_{k},\lambda_{l})}-\frac{g(\lambda_{l},\mu_{k})}{h(\lambda_{l},\mu_{k})}\mathrm{e}^{-\mathrm{i}\lambda_{l}L/2}\prod_{m=1}^{N}\frac{f(\lambda_{l},\mu_{m})}{f(\mu_{m},\lambda_{l})}\right]_{lk}\label{eq:Slavnov-overlap0}
\end{equation}
with $\boldsymbol{\mu}$} and $\boldsymbol{\lambda}$ satisfying (\ref{eq:BAeqs_L/2})
and (\ref{eq:BAeqs_L}) respectively. This result was found in \cite{Mossel-Caux}. 

The above expression is highly inconvenient for expressing the time
evolved wave-function in a multiple integral form, which is important
for passing to the thermodynamic limit. The reason is that the overlaps
are not continuous functions of the rapidities $\boldsymbol{\lambda}$
and in their original form (\ref{eq:Slavnov-overlap0}) it is not
clear how to obtain a convenient reformulation. At this point we should
recall from \cite{paper1} that, in order to express the overlaps
$\mathcal{M}(\boldsymbol{\lambda};\boldsymbol{\mu})$ as functions
of continuous variables $\boldsymbol{\lambda}$, it was crucial to
first eliminate the $L$ dependence which enters through the factors
$\mathrm{e}^{-\mathrm{i}\lambda_{l}L/2}$. We do this by replacing
these factors using the Bethe equations (\ref{eq:BAeqs_L}). Taking
the square root of both sides of the equation, we find that 
\begin{equation}
\mathrm{e}^{-\mathrm{i}\lambda_{l}L/2}=\pm\sqrt{\left(\prod_{n(\neq l)}^{N}\frac{f(\lambda_{l},\lambda_{n})}{f(\lambda_{n},\lambda_{l})}\right)^{-1}}\label{eq:sub}
\end{equation}
with either a plus or a minus sign, depending on whether in a given
solution $\boldsymbol{\lambda}$ of the Bethe equations, $Q_{l}(\boldsymbol{\lambda})$
is an even or odd integer multiple of $2\pi$ respectively. We therefore
introduce discrete indices $s_{l}$ to distinguish the two cases,
exactly as we did in the Tonks-Girardeau case \cite{paper1}. More
explicitly, we define
\begin{equation}
s_{i}\coloneqq\mathrm{e}^{-\mathrm{i}Q_{i}(\boldsymbol{\lambda})/2}=\begin{cases}
+1 & \text{if }Q_{i}(\boldsymbol{\lambda})/(2\pi)\text{ even}\\
-1 & \text{if }Q_{i}(\boldsymbol{\lambda})/(2\pi)\text{ odd}
\end{cases}\quad,\text{for }\boldsymbol{\lambda}:\text{BA}\label{eq:s_def}
\end{equation}
for any solution $\boldsymbol{\lambda}$ of the Bethe equations (\ref{eq:BAeqs_L}).
We can now redefine $\mathcal{M}(\boldsymbol{\lambda};\boldsymbol{\mu})$
to be a function of continuous rapidity variables $\boldsymbol{\lambda}$
and discrete indices $\boldsymbol{s}$. Substituting (\ref{eq:sub})
into (\ref{eq:Slavnov-overlap0}) and taking the complex conjugate\foreignlanguage{english}{
using the definitions of $g,f,h$ (\ref{eq:gfh},\ref{eq:r}) we finally
find}
\begin{equation}
\mathcal{M}_{\boldsymbol{s}}(\boldsymbol{\lambda};\boldsymbol{\mu})=\prod_{j>k}^{N}g(\lambda_{j},\lambda_{k})g(\mu_{k},\mu_{j})\prod_{j,k}^{N}h(\lambda_{k},\mu_{j})\det\left[\frac{g(\lambda_{l},\mu_{k})}{h(\lambda_{l},\mu_{k})}-s_{l}\left(\prod_{n(\neq l)}^{N}\frac{f(\lambda_{l},\lambda_{n})}{f(\lambda_{n},\lambda_{l})}\right)^{1/2}\frac{g(\mu_{k},\lambda_{l})}{h(\mu_{k},\lambda_{l})}\prod_{m=1}^{N}\frac{f(\mu_{m},\lambda_{l})}{f(\lambda_{l},\mu_{m})}\right]_{lk}\label{eq:Slavnov-overlap1}
\end{equation}

\selectlanguage{english}%
\foreignlanguage{british}{Writing $S_{N}(\boldsymbol{\mu};\boldsymbol{\lambda})$
in the alternative form (\ref{eq:Slavnov-det}) we also have}
\begin{equation}
\mathcal{M}_{\boldsymbol{s}}(\boldsymbol{\lambda};\boldsymbol{\mu})=\frac{1}{\det\left[g(\lambda_{l},\mu_{k})\right]_{lk}}\det\left[g^{2}(\lambda_{l},\mu_{k})\left(\prod_{m(\neq k)}^{N}f(\lambda_{l},\mu_{m})-s_{l}\prod_{n(\neq l)}^{N}\left(\frac{f(\lambda_{l},\lambda_{n})}{f(\lambda_{n},\lambda_{l})}\right)^{1/2}\prod_{m(\neq k)}^{N}f(\mu_{m},\lambda_{l})\right)\right]_{lk}\label{eq:Slavnov-overlap2}
\end{equation}

\selectlanguage{british}%
To lighten the notation, we now introduce the functions $\tilde{Q}_{i}(\boldsymbol{\lambda})$
and $\tilde{Q}(\lambda_{i};\boldsymbol{\mu})$ defined as follows
\begin{align}
\exp\left(-\mathrm{i}\tilde{Q}_{i}(\boldsymbol{\lambda})\right) & \coloneqq\prod_{j(\neq i)}^{N}\frac{f(\lambda_{i},\lambda_{j})}{f(\lambda_{j},\lambda_{i})}\label{eq:Q~}\\
\exp\left(-\mathrm{i}\tilde{Q}(\lambda_{i};\boldsymbol{\mu})\right) & \coloneqq-\prod_{j=1}^{N}\frac{f(\lambda_{i},\mu_{j})}{f(\mu_{j},\lambda_{i})}\label{eq:Q~2}
\end{align}
The relation of $\tilde{Q}_{i}(\boldsymbol{\lambda})$ with $Q_{i}^{(0)}(\boldsymbol{\lambda})$
and $Q_{i}(\boldsymbol{\lambda})$ defined in (\ref{eq:Q0def}) and
(\ref{eq:Qdef}) is 
\begin{align}
\exp\left(-\mathrm{i}Q_{i}^{(0)}(\boldsymbol{\lambda})\right) & =\mathrm{e}^{-\mathrm{i}\lambda_{i}L/2}\exp\left(-\mathrm{i}\tilde{Q}_{i}(\boldsymbol{\lambda})\right)\label{eq:Q0Q~}\\
\exp\left(-\mathrm{i}Q_{i}(\boldsymbol{\lambda})\right) & =\mathrm{e}^{-\mathrm{i}\lambda_{i}L}\exp\left(-\mathrm{i}\tilde{Q}_{i}(\boldsymbol{\lambda})\right)\label{eq:QQ~}
\end{align}
respectively.  Moreover, notice that
\begin{equation}
\lim_{\lambda_{i}\to\mu_{i}}\exp\left(-\mathrm{i}\tilde{Q}(\lambda_{i};\boldsymbol{\mu})\right)=\prod_{j(\neq i)}^{N}\frac{f(\mu_{i},\mu_{j})}{f(\mu_{j},\mu_{i})}=\exp\left(-\mathrm{i}\tilde{Q}_{i}(\boldsymbol{\mu})\right)\label{eq:Q~lim}
\end{equation}

Using these definitions we can write (\ref{eq:Slavnov-overlap2})
in the more compact form\textbf{}\foreignlanguage{english}{
\begin{equation}
\mathcal{M}_{\boldsymbol{s}}(\boldsymbol{\lambda};\boldsymbol{\mu})=G_{N}(\boldsymbol{\lambda},\boldsymbol{\mu})\det\left[\frac{g(\lambda_{l},\mu_{k})}{h(\lambda_{l},\mu_{k})}+s_{l}\frac{g(\mu_{k},\lambda_{l})}{h(\mu_{k},\lambda_{l})}\exp\left(-\mathrm{i}\tilde{Q}_{l}(\boldsymbol{\lambda})/2+\mathrm{i}\tilde{Q}(\lambda_{l};\boldsymbol{\mu})\right)\right]_{lk}\label{eq:Slavnov-overlap3}
\end{equation}
}

\selectlanguage{english}%
\foreignlanguage{british}{ }

\selectlanguage{british}%
Let us now focus on the analyticity properties of the above overlaps
for real rapidities $\boldsymbol{\lambda}$. From \foreignlanguage{english}{(\ref{eq:gfh})}
we notice that the function $g(\lambda,\lambda')$ has a pole at $\lambda=\lambda'$,
the function $f(\lambda,\lambda')$ has also a pole at the same point,
but the ratio $f(\lambda,\lambda')/f(\lambda',\lambda)$ does not,
and lastly the function $h(\lambda,\lambda')$ has neither poles nor
zeros for real $\lambda$. Therefore $G_{N}(\boldsymbol{\lambda},\boldsymbol{\mu})$
has no poles, while each matrix element $\left[M_{N}(\boldsymbol{\mu};\boldsymbol{\lambda})\right]_{lk}$
has a pole at $\lambda_{l}=\mu_{k}$ due to the $g$ functions. We
therefore find that the overlaps $\mathcal{M}_{\boldsymbol{s}}(\boldsymbol{\lambda};\boldsymbol{\mu})$
exhibit simple poles when any of the rapidities $\lambda_{i}$ tends
to any of the $\mu_{j}$ i.e. they exhibit $N$-dimensional poles
at $\boldsymbol{\lambda}\to\boldsymbol{\mu}$ and permutations thereof,
all of which have the same residue, since as explained in Sec. \ref{subsec:Slavnov_formula}
the Slavnov formula is symmetric under rapidity permutations. In
addition, there are singularities coming from the square root factors
$\exp(-\mathrm{i}\tilde{Q}_{l}(\boldsymbol{\lambda})/2)$. Indeed,
due to the factors $f(\lambda_{i},\lambda_{j})/f(\lambda_{j},\lambda_{i})$
in (\ref{eq:Q~}), the function $\exp(-\mathrm{i}\tilde{Q}_{l}(\boldsymbol{\lambda}))$
has poles and zeroes when $\lambda_{i}$ approaches the points $\lambda_{j}\pm{\rm i}c$
where $\lambda_{j}$ is any other rapidity. When the rapidities $\boldsymbol{\lambda}$
are in the neighbourhood of the real axes, these points are all distant
from them, lying on lines parallel to the real axes at distance equal
to $c$. However, the square roots in $\exp(-\mathrm{i}\tilde{Q}_{l}(\boldsymbol{\lambda})/2)$
give rise to branch cuts in the complex $\lambda_{i}$-plane that
start from and connect the pairs of points $\lambda_{j}\pm{\rm i}c$,
and therefore cross the real $\lambda_{i}$-axis at the positions
$\lambda_{i}=\lambda_{j}$ for each index $j$. These are all the singularities
of $\mathcal{M}_{\boldsymbol{s}}(\boldsymbol{\lambda};\boldsymbol{\mu})$.

Analogously to the non-interacting case, the poles at $\boldsymbol{\lambda}=\boldsymbol{\mu}$
and permutations are of ``kinematical'' type and they reflect the
elasticity of particle scattering in integrable models. Their presence
is a characteristic property of scalar products of Bethe states and,
together with recursion relations satisfied by their residues and
other basic requirements (permutation symmetry and decay properties
with respect to the rapidity arguments), they are sufficient to completely
determine the scalar product. In fact this is precisely the way the
Slavnov formula was derived in \cite{Slavnov}. 

The branch cut singularities, on the other hand, are a consequence
of the initial splitting of the system in two halves, which results
in the different functional form of the overlaps in the odd and even
sectors, in combination with the fact that the S-matrix (\ref{eq:S-matrix})
is negative at small rapidity differences, which is a general property
of Bethe Ansatz solvable models.

\selectlanguage{english}%

\selectlanguage{british}%

\subsection{Eigenstate summation through multivariable Cauchy's integral formula\label{subsec:Cauchy}}

The second problem we encounter is the summation over post-quench
eigenstates. The discrete allowed values of post-quench eigenstate
rapidities $\boldsymbol{\lambda}$ in (\ref{eq:int1}) are solutions
of the non-linear highly complicated Bethe equations, therefore they
are not explicitly known as in the Tonks-Girardeau case with periodic
boundary conditions. Luckily also this problem can be circumvented,
using essentially the same complex analysis trick used in the Tonks-Girardeau
case \cite{paper1}: The sum over eigenstates can be still transformed
into a (multiple) contour integral over complex rapidities by introducing
a suitable function of the rapidities that has simple poles of residue
$1/(2\pi\mathrm{i})$ precisely at the roots of Bethe equations, and
distinguishes between odd and even integer sectors for each rapidity
$\lambda_{j}$ through the corresponding discrete index $s_{j}$. 

Such a function is indeed possible to construct, despite the fact
that the Bethe roots are only implicitly known \cite{Kitanine2005a:contour,Goehmann2004:contour,Kitanine2005:contour,Slavnov:lecture-notes}.
As we explain in detail in App. \ref{app:F}, the suitable function
is 
\begin{equation}
F_{\boldsymbol{s}}(\boldsymbol{\lambda})\coloneqq\frac{1}{(4\pi)^{N}}\varrho_{N,L}(\boldsymbol{\lambda})\prod_{i=1}^{N}\frac{1}{1-s_{i}\mathrm{e}^{-\mathrm{i}Q_{i}(\boldsymbol{\lambda})/2}}\label{eq:F1b}
\end{equation}
where $\varrho_{N,L}(\boldsymbol{\lambda})$ is defined in (\ref{eq:DoS}). 

Attention must again be paid to the singularities of this function,
since we need to ensure that not only the right function has poles
at the right points, but also that it is otherwise analytic in the
region of the complex plane scanned by the contours when deformed
as required for the evaluation of the asymptotics.  We observe that
the function $F_{\boldsymbol{s}}(\boldsymbol{\lambda})$ has the same
branch cut singularities as the overlaps $\mathcal{M}_{\boldsymbol{s}}(\boldsymbol{\lambda};\boldsymbol{\mu})$,
due to the square roots $\exp(-\mathrm{i}Q_{i}(\boldsymbol{\lambda})/2)$
appearing in (\ref{eq:F1b}), which are the same as those of $\exp(-\mathrm{i}\tilde{Q}_{i}(\boldsymbol{\lambda})/2)$
in (\ref{eq:Slavnov-overlap3}). In the Tonks-Girardeau limit studied
in \cite{paper1}, these square roots were not a real problem, because
in that limit the function $\exp\left(-\mathrm{i}Q_{i}(\boldsymbol{\lambda})\right)$
reduces to $(-1)^{N-1}\exp\left(-\mathrm{i}\lambda_{i}L\right)$ which
has neither poles nor zeroes. On the contrary, in the present general
case the function $\exp(-\mathrm{i}Q_{i}(\boldsymbol{\lambda}))$,
like $\exp(-\mathrm{i}\tilde{Q}_{i}(\boldsymbol{\lambda}))$, has
poles and zeroes when $\lambda_{i}$ is at the points $\lambda_{j}\pm{\rm i}c$,
due to the factors $f(\lambda_{i},\lambda_{j})/f(\lambda_{j},\lambda_{i})$,
and the square root in $\exp(-\mathrm{i}Q_{i}(\boldsymbol{\lambda})/2)$
gives rise to branch cuts crossing the real axes at the positions
$\lambda_{i}=\lambda_{j}$, i.e. whenever two rapidities are equal
to each other.

Even though these singularities can be removed by a redefinition of
$F_{\boldsymbol{s}}(\boldsymbol{\lambda})$ such that the branch cuts
are diverted to imaginary infinity instead of crossing the real axis
(which can be done by introducing $-1$ factors in the arguments of
the square roots, compensated by an equal number of imaginary unit
factors outside of them), such a change would not really solve the
problem. First, the cost of this change would be a modification of
the signs that the corrected function $\mathrm{e}^{-\mathrm{i}Q_{i}(\boldsymbol{\lambda})/2}$
has at the Bethe roots, which would be different from the $s_{i}$
indices, introducing an additional complication. On the other hand,
while the function $F_{\boldsymbol{s}}(\boldsymbol{\lambda})$ can
be redefined for convenience, the overlaps $\mathcal{M}_{\boldsymbol{s}}(\boldsymbol{\lambda};\boldsymbol{\mu})$
would still suffer from the same branch cut singularities, therefore
the actual problem would be still present. 

Therefore, we need to re-evaluate the analyticity requirements and
ask ourselves if they can be relaxed at no cost for the purposes of
our subsequent analysis. As we will now see this is indeed possible.
First, let us recall that the summation over Bethe states in (\ref{eq:int1})
refers to ordered rapidities and passing from sum to integral is also
restricted by construction to the domain $\lambda_{1}<\lambda_{2}<\lambda_{3}<\dots<\lambda_{N}$.
Since the summand (and corresponding integrand) is symmetric under
permutation of the rapidities, it is possible to raise the ordering
constraint and extend the rapidity domain to the entire real axis
dividing by $N!$, which is generally convenient but not necessary.
In the present case, we note that we do not aim at deforming the integration
contours away from the real rapidity axis, since for the asymptotic
analysis we need to focus on the contribution of the kinematical poles,
which also lie on the real axis like the Bethe roots themselves. Therefore,
it is not required to extend the integration to the entire real rapidity
axes and there is no real obstacle to contour deformation due to the
branch cut singularities of the integrand at the boundaries of this
domain $\lambda_{i}=\lambda_{i+1}$ for any index $i$, as soon as
we can keep the boundaries of the integration region fixed.

Recall now the multivariable version of the residue theorem \cite{paper1,Slavnov:lecture-notes},
which states that, given a function $F(\boldsymbol{\lambda})=1/\prod_{i=1}^{N}f_{i}(\boldsymbol{\lambda})$
such that $f_{i}(\boldsymbol{\lambda})$ has simple zeroes at the
collection of points $\{\boldsymbol{\lambda}^{*}\}$, a multi-dimensional
contour $\boldsymbol{C}=C_{1}\times C_{2}\times\dots\times C_{N}$
encircling these points, and a function $g(\boldsymbol{\lambda})$
of the rapidities $\boldsymbol{\lambda}$ that is analytic inside
the contour $\boldsymbol{C}$ except at the points $\{\boldsymbol{\lambda}_{p}\}$
which are all different from $\{\boldsymbol{\lambda}^{*}\}$, we have
 
\begin{equation}
\oint_{\boldsymbol{C}}\frac{{\rm d}^{N}\boldsymbol{\lambda}}{(2\pi{\rm i})^{N}}\,F(\boldsymbol{\lambda})g(\boldsymbol{\lambda})=\sum_{\boldsymbol{\lambda}^{*}}\frac{g(\boldsymbol{\lambda}^{*})}{\left.\det\left(\frac{\partial f_{i}}{\partial\lambda_{j}}\right)\right|_{\boldsymbol{\lambda}=\boldsymbol{\lambda}^{*}}}+\sum_{\boldsymbol{\lambda}_{p}}F(\boldsymbol{\lambda}_{p})\underset{\boldsymbol{\lambda}=\boldsymbol{\lambda}_{p}}{\text{Res}}g(\boldsymbol{\lambda})\label{eq:trick}
\end{equation}
This allows us to write the sum over Bethe states as
\[
\sum_{\boldsymbol{\lambda}:\text{ BA}}\dots\,=\frac{1}{N!}\sum_{\boldsymbol{s}}\oint_{\boldsymbol{C}}{\rm d}^{N}\boldsymbol{\lambda}\,F_{\boldsymbol{s}}(\boldsymbol{\lambda})\dots
\]
where the kinematical poles should be excluded from the contour $\boldsymbol{C}$.

Using the above defined function (\ref{eq:F1b}) and substituting
(\ref{eq:Slavnov-overlap3}) and (\ref{eq:norm}), the eigenstate
sum in (\ref{eq:int1}) can be written in integral form as
\begin{align}
 & \mathrm{e}^{+\mathrm{i}Px-\mathrm{i}Ht}|\Phi_{0}(\boldsymbol{\mu})\rangle\nonumber \\
 & =\sum_{\boldsymbol{\lambda}:\text{ BA}}\mathrm{e}^{+\mathrm{i}P(\boldsymbol{\lambda})x-\mathrm{i}E(\boldsymbol{\lambda})t}\frac{\mathcal{M}(\boldsymbol{\lambda};\boldsymbol{\mu})}{\mathcal{N}(\boldsymbol{\lambda})}|\Phi(\boldsymbol{\lambda})\rangle\nonumber \\
 & =\sum_{\boldsymbol{s}}\oint_{\boldsymbol{C}}{\rm d}^{N}\boldsymbol{\lambda}\,F_{\boldsymbol{s}}(\boldsymbol{\lambda})\mathrm{e}^{+\mathrm{i}P(\boldsymbol{\lambda})x-\mathrm{i}E(\boldsymbol{\lambda})t}\frac{\mathcal{M}_{\boldsymbol{s}}(\boldsymbol{\lambda};\boldsymbol{\mu})}{\mathcal{N}(\boldsymbol{\lambda})}|\Phi(\boldsymbol{\lambda})\rangle\nonumber \\
 & =\oint_{\boldsymbol{C}}\frac{{\rm d}^{N}\boldsymbol{\lambda}}{(4\pi)^{N}}\,\mathrm{e}^{+\mathrm{i}P(\boldsymbol{\lambda})x-\mathrm{i}E(\boldsymbol{\lambda})t}\frac{G_{N}(\boldsymbol{\lambda},\boldsymbol{\mu})}{c^{N}\left(\prod_{m\neq\ell}f(\lambda_{m},\lambda_{\ell})\right)}\nonumber \\
 & \qquad\times\sum_{\boldsymbol{s}}\prod_{i=1}^{N}\frac{1}{1-s_{i}\mathrm{e}^{-\mathrm{i}Q_{i}(\boldsymbol{\lambda})/2}}\det\left[\frac{g(\lambda_{l},\mu_{k})}{h(\lambda_{l},\mu_{k})}+s_{l}\frac{g(\mu_{k},\lambda_{l})}{h(\mu_{k},\lambda_{l})}\mathrm{e}^{-\mathrm{i}\tilde{Q}_{l}(\boldsymbol{\lambda})/2+\mathrm{i}\tilde{Q}(\lambda_{l};\boldsymbol{\mu})}\right]_{lk}|\Phi(\boldsymbol{\lambda})\rangle\label{eq:int2}
\end{align}
where the multi-contour $\boldsymbol{C}$ in the above integral
is defined as 
\begin{align}
\boldsymbol{C} & =C_{1}\times C_{2}\times\dots\times C_{N}\label{eq:contour}\\
C_{i} & =C_{i}^{(r)}-C_{i}^{(p)}\nonumber \\
C_{i}^{(r)} & =(\lambda_{i-1}+\epsilon-\mathrm{i}\epsilon,\lambda_{i+1}-\epsilon-\mathrm{i}\epsilon,\lambda_{i+1}-\epsilon+\mathrm{i}\epsilon,\lambda_{i-1}+\epsilon+\mathrm{i}\epsilon,\lambda_{i-1}+\epsilon-\mathrm{i}\epsilon)\nonumber \\
C_{i}^{(p)} & =C(\mu_{i},\eta)\nonumber 
\end{align}
In more detail, each of the contours $C_{i}$ should enclose the interval
$(\lambda_{i-1},\lambda_{i+1})$ of the real rapidity axis, where
all Bethe roots lie in the repulsive case, but it should exclude the
kinematical pole of the overlap $\mathcal{M}_{\boldsymbol{s}}(\boldsymbol{\lambda};\boldsymbol{\mu})$
at $\lambda_{i}\to\mu_{i}$. This is achieved by choosing $C_{i}$
to consist of a thing rectangle $C_{i}^{(r)}$ composed by two straight
lines one just above and one just below this real axis interval, and
subtracting a small circle $C_{i}^{(p)}$ around the the value $\mu_{i}$.
 Note that this pole is between the Bethe roots and does not coincide
with any of them in general. This is because the roots of the half-system
Bethe equations are not roots of the full-system Bethe equations,
since otherwise by dividing (\ref{eq:BAeqs_L/2}) and (\ref{eq:BAeqs_L})
we would have that $\mathrm{e}^{\mathrm{i}\mu_{i}L/2}=1$ i.e. the
corresponding rapidities are integer multiples of $4\pi/L$. This
means that the occasion of coinciding pre- and post-quench momenta
for even eigenstates that occurs in the Tonks-Girardeau case for odd
$N$ \cite{paper1} is exceptional and not present in the interacting
case. Also note that all other constituents of the integrand in (\ref{eq:int2})
i.e. the functions $\mathrm{e}^{+\mathrm{i}P(\boldsymbol{\lambda})x-\mathrm{i}E(\boldsymbol{\lambda})t}$,
$\mathcal{N}^{-1}(\boldsymbol{\lambda})$ and the state $|\Phi(\boldsymbol{\lambda})\rangle$
itself, are analytic functions for real rapidities and do not introduce
any other singularities.

In addition, the summation over the discrete indices $\boldsymbol{s}$
can be perfomed at this step. By absorbing the product of (\ref{eq:F1b})
and the sum over $\boldsymbol{s}$ into the determinant of (\ref{eq:int2}),
we can write the expression of the last line of (\ref{eq:int2}) as

\begin{align}
 & \sum_{\boldsymbol{s}}\prod_{i=1}^{N}\frac{1}{1-s_{i}\mathrm{e}^{-\mathrm{i}Q_{i}(\boldsymbol{\lambda})/2}}\det\left[\frac{g(\lambda_{l},\mu_{k})}{h(\lambda_{l},\mu_{k})}+s_{l}\frac{g(\mu_{k},\lambda_{l})}{h(\mu_{k},\lambda_{l})}\mathrm{e}^{-\mathrm{i}\tilde{Q}_{l}(\boldsymbol{\lambda})/2+\mathrm{i}\tilde{Q}(\lambda_{l};\boldsymbol{\mu})}\right]_{lk}\nonumber \\
 & =\det\left[\sum_{s}\frac{\frac{g(\lambda_{l},\mu_{k})}{h(\lambda_{l},\mu_{k})}+s\frac{g(\mu_{k},\lambda_{l})}{h(\mu_{k},\lambda_{l})}\mathrm{e}^{-\mathrm{i}\tilde{Q}_{l}(\boldsymbol{\lambda})/2+\mathrm{i}\tilde{Q}(\lambda_{l};\boldsymbol{\mu})}}{1-s\mathrm{e}^{-\mathrm{i}Q_{l}(\boldsymbol{\lambda})/2}}\right]_{lk}\nonumber \\
 & =\det\left[2\frac{\frac{g(\lambda_{l},\mu_{k})}{h(\lambda_{l},\mu_{k})}+\frac{g(\mu_{k},\lambda_{l})}{h(\mu_{k},\lambda_{l})}\mathrm{e}^{-\mathrm{i}Q_{l}(\boldsymbol{\lambda})/2-\mathrm{i}\tilde{Q}_{l}(\boldsymbol{\lambda})/2+\mathrm{i}\tilde{Q}(\lambda_{l};\boldsymbol{\mu})}}{1-\mathrm{e}^{-\mathrm{i}Q_{l}(\boldsymbol{\lambda})}}\right]_{lk}\nonumber \\
 & =2^{N}\prod_{i=1}^{N}\frac{1}{1-\mathrm{e}^{-\mathrm{i}Q_{i}(\boldsymbol{\lambda})}}\det\left[\frac{g(\lambda_{l},\mu_{k})}{h(\lambda_{l},\mu_{k})}+\frac{g(\mu_{k},\lambda_{l})}{h(\mu_{k},\lambda_{l})}\mathrm{e}^{-\mathrm{i}Q_{l}(\boldsymbol{\lambda})/2-\mathrm{i}\tilde{Q}_{l}(\boldsymbol{\lambda})/2+\mathrm{i}\tilde{Q}(\lambda_{l};\boldsymbol{\mu})}\right]_{lk}\label{eq:sumFM}
\end{align}
Substituting back to (\ref{eq:int2}), we obtain the alternative
formula 
\begin{align}
 & \mathrm{e}^{+\mathrm{i}Px-\mathrm{i}Ht}|\Phi_{0}(\boldsymbol{\mu})\rangle\nonumber \\
 & =\oint_{\boldsymbol{C}}\frac{{\rm d}^{N}\boldsymbol{\lambda}}{(2\pi)^{N}}\,\mathrm{e}^{+\mathrm{i}P(\boldsymbol{\lambda})x-\mathrm{i}E(\boldsymbol{\lambda})t}\frac{G_{N}(\boldsymbol{\lambda},\boldsymbol{\mu})}{c^{N}\left(\prod_{m\neq\ell}f(\lambda_{m},\lambda_{\ell})\right)}\nonumber \\
 & \qquad\times\prod_{i=1}^{N}\frac{1}{1-\mathrm{e}^{-\mathrm{i}Q_{i}(\boldsymbol{\lambda})}}\det\left[\frac{g(\lambda_{l},\mu_{k})}{h(\lambda_{l},\mu_{k})}+\frac{g(\mu_{k},\lambda_{l})}{h(\mu_{k},\lambda_{l})}\mathrm{e}^{-\mathrm{i}Q_{l}(\boldsymbol{\lambda})/2-\mathrm{i}\tilde{Q}_{l}(\boldsymbol{\lambda})/2+\mathrm{i}\tilde{Q}(\lambda_{l};\boldsymbol{\mu})}\right]_{lk}|\Phi(\boldsymbol{\lambda})\rangle\label{eq:int2b}
\end{align}
Note that in passing from (\ref{eq:int2}) to (\ref{eq:int2b}) using
(\ref{eq:sumFM}), we have implicitly used that the eigenstates $|\Phi(\boldsymbol{\lambda})\rangle$
are essentially independent of the indices $\boldsymbol{s}$. This
is true as long as we are interested in the dynamics of the system
in the bulk in the thermodynamic limit. Indeed, the coordinate space
wave-functions of the Bethe eigenstates (\ref{eq:coord_BA_wf}) are
continuous functions of the rapidities $\boldsymbol{\lambda}$ in
the bulk of the system, independent of the quantisation conditions
(\ref{eq:BAeqs}), therefore, independent of whether they correspond
to odd or even quantum numbers. The dependence on the indices $\boldsymbol{s}$
is only evident close to the boundaries, which move away to infinity
in the thermodynamic limit. 

Eqs. (\ref{eq:int2}) and (\ref{eq:int2b}) together with the definition
(\ref{eq:contour}) of the multi-contour $\boldsymbol{C}$ are our
main results for the time evolved many-body wave-function expressed
in a multiple integral representation. The integrands in these expressions
involve determinants originating from the Slavnov formula with adjustments
and manipulation specific to the geometric quench problem.

\section{Multivariable Kinematical Pole Residue\label{sec:pole}}

Having identified the analyticity properties and the location of singularities
of the integrands of (\ref{eq:int2}) and (\ref{eq:int2b}) paves
the way for the derivation of the asymptotics in the combined thermodynamic
and large distance and time limit. As an additional step towards this
direction, we will now focus on the contribution of the multivariable
kinematical pole at $\boldsymbol{\lambda}\to\boldsymbol{\mu}$ to
the integral, calculating its residue. As anticipated based on the
Tonks-Girardeau calculation \cite{paper1}, this is expected to give
the only non-vanishing contribution in the thermodynamic and large
time and distance limit. 

Explicitly, we split the integral in (\ref{eq:int2}) into two parts,
one corresponding to the $N$-dimensional kinematical pole residue
and the remainder $\mathcal{R}$ consisting of all other contributions,
i.e. all cross terms corresponding to products of lower-order pole
residues and integrals 
\begin{align}
\mathrm{e}^{+\mathrm{i}Px-\mathrm{i}Ht}|\Phi_{0}(\boldsymbol{\mu})\rangle & =(-2\pi\mathrm{i})^{N}\sum_{\boldsymbol{s}}F_{\boldsymbol{s}}(\boldsymbol{\mu})\mathrm{e}^{+\mathrm{i}P(\boldsymbol{\mu})x-\mathrm{i}E(\boldsymbol{\mu})t}\frac{\underset{\boldsymbol{\lambda}=\boldsymbol{\mu}}{\text{Res}}\mathcal{M}_{\boldsymbol{s}}(\boldsymbol{\lambda};\boldsymbol{\mu})}{\mathcal{N}(\boldsymbol{\mu})}|\Phi(\boldsymbol{\mu})\rangle+\mathcal{R}\label{eq:int-res-1}
\end{align}
Note that the presence of the minus sign in $(-2\pi\mathrm{i})^{N}$
is due to the pole contribution being subtracted from each of the
rapidity integrals. 

We start by evaluating $F_{\boldsymbol{s}}(\boldsymbol{\lambda})$
at $\boldsymbol{\lambda}=\boldsymbol{\mu}$. Since the rapidities
$\boldsymbol{\mu}$ satisfy the Bethe equations in the half system
(\ref{eq:BAeqs_L/2}), we have
\begin{align}
\exp\left(-\mathrm{i}Q_{j}(\boldsymbol{\mu})\right) & =\exp(-\mathrm{i}\mu_{j}L)\exp\left(-\mathrm{i}\tilde{Q}_{j}(\boldsymbol{\mu})\right)\nonumber \\
 & =\exp(-\mathrm{i}\mu_{j}L/2)\exp\left(-\mathrm{i}Q_{j}^{(0)}(\boldsymbol{\mu})\right)\nonumber \\
 & =\exp(-\mathrm{i}\mu_{j}L/2)\label{eq:expiQm}
\end{align}
from which we find
\begin{align*}
\exp\left(-\mathrm{i}Q_{j}(\boldsymbol{\mu})/2\right) & =\rho_{j}\exp(-\mathrm{i}\mu_{j}L/4)
\end{align*}
\foreignlanguage{english}{where, similarly to $s_{i}$, we have introduced
the sign $\rho_{i}\coloneqq\mathrm{e}^{-\mathrm{i}Q_{i}^{(0)}(\boldsymbol{\mu})/2}=\pm1$
corresponding to the $i$-th rapidity of the set $\boldsymbol{\mu}$
considered as a solution of the Bethe equations for the half system.}
Therefore, from (\ref{eq:F1b}) we obtain

\begin{align}
F_{\boldsymbol{s}}(\boldsymbol{\mu}) & =\frac{1}{(4\pi)^{N}}\left(\prod_{i=1}^{N}\frac{1}{1-s_{i}\mathrm{e}^{-\mathrm{i}Q_{i}(\boldsymbol{\mu})/2}}\right)\varrho_{N,L}(\boldsymbol{\mu})\nonumber \\
 & =\frac{1}{(4\pi)^{N}}\left(\prod_{i=1}^{N}\frac{1}{1-s_{i}\rho_{i}\mathrm{e}^{-\mathrm{i}\mu_{i}L/4}}\right)\varrho_{N,L}(\boldsymbol{\mu})\label{eq:F_int1-2}
\end{align}

\selectlanguage{english}%

\selectlanguage{british}%
Next we evaluate the residue of the overlaps $\mathcal{M}_{\boldsymbol{s}}(\boldsymbol{\lambda};\boldsymbol{\mu})$
at $\boldsymbol{\lambda}=\boldsymbol{\mu}$ from (\ref{eq:Slavnov-overlap1}).
As mentioned earlier, the $N$-dimensional pole of $\mathcal{M}_{\boldsymbol{s}}(\boldsymbol{\lambda};\boldsymbol{\mu})$
at this point is due to the presence of the functions \foreignlanguage{english}{$g(\mu,\lambda)=\mathrm{i}c/(\mu-\lambda)$
in the matrix $M_{N}(\boldsymbol{\mu};\boldsymbol{\lambda})$.} By
expanding the $\det\left[M_{N}(\boldsymbol{\mu};\boldsymbol{\lambda})\right]_{lk}$
as a sum over permutations of products of matrix elements, we see
that there is exactly one term that contains all ($N$ in number)
singular factors \foreignlanguage{english}{$g(\mu_{k},\lambda_{k})$}
and it is the only one that contributes to the multi-variable residue.
Each of the other terms has a $N$-dimensional pole with the same
residue at $\boldsymbol{\lambda}$ equal to one of the permutations
of $\boldsymbol{\mu}$. Recall that in the calculation of the norm
$\mathcal{N}(\boldsymbol{\mu})$ outlined in Sec. \ref{subsec:Gaudin-formula},
the poles of the $g$ functions\foreignlanguage{english}{ are cancelled
by zeros when we set }$\boldsymbol{\lambda}\to\boldsymbol{\mu}$\foreignlanguage{english}{
and substitute the Bethe equations. On the contrary, this is not the
case here since the fact that $\boldsymbol{\lambda}$ and $\boldsymbol{\mu}$
satisfy different Bethe equations results in the different functional
form of the determinant appearing in (\ref{eq:Slavnov-overlap3})
when considered as a function of continuous rapidities $\boldsymbol{\lambda}$.
More specifically, in the present case we have}

\selectlanguage{english}%
\begin{align*}
 & \underset{\boldsymbol{\lambda}=\boldsymbol{\mu}}{\text{Res}}\det\left[\frac{g(\lambda_{l},\mu_{k})}{h(\lambda_{l},\mu_{k})}+s_{l}\frac{g(\mu_{k},\lambda_{l})}{h(\mu_{k},\lambda_{l})}\mathrm{e}^{-\mathrm{i}\tilde{Q}_{l}(\boldsymbol{\lambda})/2+\mathrm{i}\tilde{Q}(\lambda_{l};\boldsymbol{\mu})}\right]_{lk}\\
 & =\underset{\boldsymbol{\lambda}=\boldsymbol{\mu}}{\text{Res}}\prod_{k=1}^{N}\left[\frac{g(\lambda_{k},\mu_{k})}{h(\lambda_{k},\mu_{k})}+s_{k}\frac{g(\mu_{k},\lambda_{k})}{h(\mu_{k},\lambda_{k})}\mathrm{e}^{-\mathrm{i}\tilde{Q}_{k}(\boldsymbol{\lambda})/2+\mathrm{i}\tilde{Q}(\lambda_{k};\boldsymbol{\mu})}\right]\\
 & =\prod_{k=1}^{N}\frac{1}{h(\mu_{k},\mu_{k})}\left[1-s_{k}\mathrm{e}^{-\mathrm{i}\tilde{Q}_{k}(\boldsymbol{\mu})/2}\lim_{\lambda_{k}\to\mu_{k}}\mathrm{e}^{\mathrm{i}\tilde{Q}(\lambda_{k};\boldsymbol{\mu})}\right]\underset{\lambda_{k}=\mu_{k}}{\text{Res}}\,g(\lambda_{k},\mu_{k})\\
 & =\left(\mathrm{i}c\right)^{N}\prod_{k=1}^{N}\left[1-s_{k}\mathrm{e}^{-\mathrm{i}\tilde{Q}_{k}(\boldsymbol{\mu})/2+\mathrm{i}\tilde{Q}_{k}(\boldsymbol{\mu})}\right]
\end{align*}
In the third line we used that $g(\mu,\lambda)=-g(\lambda,\mu)$,
while in the last line we used (\ref{eq:Q~lim}) and replaced $\underset{\lambda=\mu}{\text{Res}}\,g(\lambda,\mu)=\mathrm{i}c$
and $h(\mu,\mu)=1$. Using once again the \foreignlanguage{british}{half
system Bethe equations (\ref{eq:BAeqs_L/2}) from which we obtained
(\ref{eq:expiQm}), we have 
\[
\mathrm{e}^{+\mathrm{i}\tilde{Q}_{i}(\boldsymbol{\mu})}=\mathrm{e}^{-\mathrm{i}\mu_{i}L/2}
\]
and 
\[
\mathrm{e}^{+\mathrm{i}\tilde{Q}_{i}(\boldsymbol{\mu})/2}=\rho_{i}\mathrm{e}^{-\mathrm{i}\mu_{i}L/4}
\]
so that we can rewrite the last result as}
\begin{align*}
 & \underset{\boldsymbol{\lambda}=\boldsymbol{\mu}}{\text{Res}}\det\left[\frac{g(\lambda_{l},\mu_{k})}{h(\lambda_{l},\mu_{k})}+s_{l}\frac{g(\mu_{k},\lambda_{l})}{h(\mu_{k},\lambda_{l})}\mathrm{e}^{-\mathrm{i}\tilde{Q}_{l}(\boldsymbol{\lambda})/2+\mathrm{i}\tilde{Q}(\lambda_{l};\boldsymbol{\mu})}\right]_{lk}\\
 & =\left(\mathrm{i}c\right)^{N}\prod_{k=1}^{N}\left(1-s_{k}\mathrm{e}^{+\mathrm{i}\tilde{Q}_{k}(\boldsymbol{\mu})/2}\right)\\
 & =\left(\mathrm{i}c\right)^{N}\prod_{k=1}^{N}\left(1-s_{k}\rho_{k}\mathrm{e}^{-\mathrm{i}\mu_{k}L/4}\right)
\end{align*}
Finally, using also (\ref{eq:Gdiag}) for $G_{N}(\boldsymbol{\mu},\boldsymbol{\mu})$,
we obtain the following result for the kinematical pole residue of
the overlaps

\selectlanguage{british}%
\begin{equation}
\underset{\boldsymbol{\lambda}=\boldsymbol{\mu}}{\text{Res}}\mathcal{M}_{\boldsymbol{s}}(\boldsymbol{\lambda};\boldsymbol{\mu})=\left(\mathrm{i}c\right)^{N}\prod_{i\neq j}^{N}f(\mu_{i},\mu_{j})\prod_{k=1}^{N}\left(1-s_{k}\rho_{k}\mathrm{e}^{-\mathrm{i}\mu_{k}L/4}\right)\label{eq:M_res-1}
\end{equation}

Substituting (\ref{eq:F_int1-2}), (\ref{eq:M_res-1}) and the norm
formula (\ref{eq:norm}), we obtain\foreignlanguage{english}{
\begin{align}
F_{\boldsymbol{s}}(\boldsymbol{\mu})\frac{\underset{\boldsymbol{\lambda}=\boldsymbol{\mu}}{\text{Res}}\mathcal{M}_{\boldsymbol{s}}(\boldsymbol{\lambda};\boldsymbol{\mu})}{\mathcal{N}(\boldsymbol{\mu})} & =\left(\frac{\mathrm{i}}{4\pi}\right)^{N}\prod_{k=1}^{N}\left(\frac{1-s_{k}\rho_{k}\mathrm{e}^{-\mathrm{i}\mu_{k}L/4}}{1-s_{k}\rho_{k}\mathrm{e}^{-\mathrm{i}\mu_{k}L/4}}\right)=\left(\frac{\mathrm{i}}{4\pi}\right)^{N}\label{eq:FMN}
\end{align}
}We notice that the value of the residue is the same \foreignlanguage{english}{independently
of the signs $s_{k}$ and $\rho_{k}$}. Using the last result, we
finally conclude that the contribution of the kinematical pole residue
in (\ref{eq:int-res-1}) is
\begin{align}
(-2\pi\mathrm{i})^{N}\sum_{\boldsymbol{s}}F_{\boldsymbol{s}}(\boldsymbol{\mu})\mathrm{e}^{+\mathrm{i}P(\boldsymbol{\mu})x-\mathrm{i}E(\boldsymbol{\mu})t}\frac{\underset{\boldsymbol{\lambda}=\boldsymbol{\mu}}{\text{Res}}\mathcal{M}_{\boldsymbol{s}}(\boldsymbol{\lambda};\boldsymbol{\mu})}{\mathcal{N}(\boldsymbol{\mu})}|\Phi(\boldsymbol{\mu})\rangle & =\mathrm{e}^{+\mathrm{i}P(\boldsymbol{\mu})x-\mathrm{i}E(\boldsymbol{\mu})t}|\Phi(\boldsymbol{\mu})\rangle\label{eq:KP_Res}
\end{align}

Compared to the long and complicated form of intermediate formulas,
this is a surprisingly simple and elegant result. Note especially
that, despite the necessary step of decomposition into odd and even
sectors, the final formula for the residue does not contain any sign
of this. It can be readily verified that the above result is consistent 
with that of \cite{paper1} for the Tonks-Girardeau limit.

\selectlanguage{english}%

\selectlanguage{british}%

\section{Discussion\label{sec:Discussion}}

In this work we have derived a multiple integral representation for
the time evolved wave-function after a geometric quench in the Lieb-Liniger
model for arbitrary interaction $c>0$. We have also discussed the
analyticity properties of the integrand, identified its singularities
and calculated the kinematical pole residue. These results serve as
a first step towards the derivation of the asymptotics of observables
in the thermodynamic and large distance and time limit, and therefore
the complete characterisation of the emergent NESS. 

The formulas derived here have been verified numerically for two and
three particles. Even though partial, these tests are useful, especially since the formulas
involve alternating signs and all steps of the calculation are very
sensitive to sign errors. More specifically, we have verified all
of the alternative forms (\ref{eq:Slavnov-overlap1}), (\ref{eq:Slavnov-overlap2})
and (\ref{eq:Slavnov-overlap3}) of the initial state overlaps (\ref{eq:Slavnov-overlap0}).
We have also verified that the function (\ref{eq:F1b}) is the correct
function for rewriting the sum over Bethe states as a multiple integral.
Moreover, we have verified the equivalence between the original sum
over Bethe states (\ref{eq:int1}) and each of the multiple integral
formulas (\ref{eq:int2}) and (\ref{eq:int2b}). To this end, we considered
closed and finite complex plane contours encircling several Bethe
roots, evaluated the resulting integrals numerically and compared
them with the sums over the specified encircled Bethe roots. This
test was performed both in the case of the kinematical pole being
located inside and outside of the chosen contour. Lastly, we have
verified the value of the kinematical pole residue (\ref{eq:KP_Res}),
especially the independence of (\ref{eq:FMN}) from the \foreignlanguage{english}{signs
$\boldsymbol{s}$ and $\boldsymbol{\rho}$ and the pre-quench rapidities
$\boldsymbol{\mu}$, both by numerical integration along small contours encircling them 
and by numerical evaluation of the limit formula.}

As discussed in the introduction, a key point in the derivation of
the asymptotics of observables using a multiple integral representation
is whether the integrand is analytic to an extent sufficient to allow
for the necessary deformations of the integration contours. In the
case of the geometric quench studied here we observe that, owing to
the form of the initial state overlaps, the integrand
exhibits inevitable singularities (branch cuts) that obstruct extensive
contour deformations. Nevertheless, the kinematical poles of the integrand
that are expected to contain all information about the asymptotics
we are interested in lie infinitesimally close to the original contours,
specifically, within the locus of the Bethe roots. As a result, the
necessary contour deformations are only infinitesimal and therefore
are not obstructed by the presence of the branch cut singularities.
Furthermore, by calculating the kinematical pole residue of the integrand,
we demonstrated that it is completely independent of the fine details
of the initial state overlaps. This is an indication that, despite
the complicated form of the initial state overlaps, detailed information
about them is irrelevant in the thermodynamic and large distance and
time limit. This suggests that different initial states that correspond
to the same rapidity density remain close to each other
also during the time evolution and result in the same asymptotic behaviour. 
This observation is consistent with general statistical physics arguments
that are the basis of the Quench Action approach.

\appendix

\section{Construction of the meromorphic function with poles at even/odd Bethe
roots\label{app:F}}

An essential step in our calculation is the transformation of the
sum over Bethe states into a multiple integral, which can be done using a
meromorphic function $F_{\boldsymbol{s}}(\boldsymbol{\lambda})$ having
poles of equal residue at the Bethe roots and selecting all those
corresponding to the sector with parity indices $\boldsymbol{s}$.
In this appendix we discuss in detail how the suitable function can
be heuristically constructed. 

From the Bethe equations (BA) in the full length system (\ref{eq:BAeqs_L}),
we can see that a simple function having poles at all Bethe roots
is 
\[
\prod_{i=1}^{N}\frac{1}{\exp\left(\mathrm{i}Q_{i}(\boldsymbol{\lambda})\right)-1}
\]
To ensure that the residue of the required function at each of these
poles equals $1/(2\pi\mathrm{i})$, we should multiply with $\det\left(\frac{\partial}{\partial\lambda_{j}}\mathrm{e}^{\mathrm{i}Q_{i}(\boldsymbol{\lambda})/2}\right)$,
therefore obtaining 
\begin{align}
 & \frac{1}{(2\pi\mathrm{i})^{N}}\prod_{i=1}^{N}\left(\frac{1}{\exp\left(\mathrm{i}Q_{i}(\boldsymbol{\lambda})\right)-1}\right)\det\left(\frac{\partial}{\partial\lambda_{j}}\mathrm{e}^{\mathrm{i}Q_{i}(\boldsymbol{\lambda})}\right)\nonumber \\
 & =\frac{1}{(2\pi)^{N}}\prod_{i=1}^{N}\left(\frac{1}{1-\mathrm{e}^{-\mathrm{i}Q_{i}(\boldsymbol{\lambda})}}\right)\det\left(\frac{\partial Q_{i}(\boldsymbol{\lambda})}{\partial\lambda_{j}}\right)\label{eq:F0}
\end{align}
However, we also need to distinguish the roots between odd and even
integer sectors for each rapidity. Writing the fraction appearing
in the product of the last expression as
\begin{align*}
\frac{1}{1-\mathrm{e}^{-\mathrm{i}Q_{i}(\boldsymbol{\lambda})}} & =\frac{1}{\left(1-\mathrm{e}^{-\mathrm{i}Q_{i}(\boldsymbol{\lambda})/2}\right)\left(1+\mathrm{e}^{-\mathrm{i}Q_{i}(\boldsymbol{\lambda})/2}\right)}
\end{align*}
we see that the first factor corresponds to the even sector and the
second to the odd. The last expression can also be written as 
\begin{align*}
 & \frac{1}{\left(1-\mathrm{e}^{-\mathrm{i}Q_{i}(\boldsymbol{\lambda})/2}\right)\left(1+\mathrm{e}^{-\mathrm{i}Q_{i}(\boldsymbol{\lambda})/2}\right)}\\
 & =\frac{1}{2}\left(\frac{1}{1-\mathrm{e}^{-\mathrm{i}Q_{i}(\boldsymbol{\lambda})/2}}+\frac{1}{1+\mathrm{e}^{-\mathrm{i}Q_{i}(\boldsymbol{\lambda})/2}}\right)\\
 & =\frac{1}{2}\sum_{s=\pm1}\frac{1}{1-s\mathrm{e}^{-\mathrm{i}Q_{i}(\boldsymbol{\lambda})/2}}
\end{align*}
Therefore the function (\ref{eq:F0}) that has poles at all Bethe
roots can be equivalently written as 
\begin{align*}
F(\boldsymbol{\lambda}) & \coloneqq\frac{1}{(4\pi)^{N}}\prod_{i=1}^{N}\sum_{s_{i}=\pm1}\frac{1}{1-s_{i}\mathrm{e}^{-\mathrm{i}Q_{i}(\boldsymbol{\lambda})/2}}\det\left(\frac{\partial Q_{i}(\boldsymbol{\lambda})}{\partial\lambda_{j}}\right)\\
 & =\sum_{\boldsymbol{s}}\frac{1}{(4\pi)^{N}}\prod_{i=1}^{N}\frac{1}{1-s_{i}\mathrm{e}^{-\mathrm{i}Q_{i}(\boldsymbol{\lambda})/2}}\varrho_{N,L}(\boldsymbol{\lambda})
\end{align*}
where we also used (\ref{eq:DoS}). From the last expression it is
clear that the right function that selects the Bethe roots belonging
to the sector corresponding to a given vector of discrete indices
$\boldsymbol{s}$ is 
\begin{equation}
F_{\boldsymbol{s}}(\boldsymbol{\lambda})\coloneqq\frac{1}{(4\pi)^{N}}\varrho_{N,L}(\boldsymbol{\lambda})\prod_{i=1}^{N}\frac{1}{1-s_{i}\mathrm{e}^{-\mathrm{i}Q_{i}(\boldsymbol{\lambda})/2}}\label{eq:F1}
\end{equation}
which is the function used in the main text. Summing the roots of
all possible sectors, we recover the entire set of Bethe roots, $F(\boldsymbol{\lambda})=\sum_{\boldsymbol{s}}F_{\boldsymbol{s}}(\boldsymbol{\lambda})$. 

\acknowledgements
A preliminary version of this work was presented at the conference ``\emph{Correlation functions of quantum integrable systems and beyond}'' (ENS-Lyon, 2017). I would like to thank Karol Kozlowski, Jean-S\'ebastien Caux, Gyorgy Feh\'er, Bal\'azs Pozsgay, Arthur Hutsalyuk and Yuan Miao for useful discussions. This work was supported by the Slovenian Research Agency (ARRS) under grant QTE (N1-0109) and by the ERC Advanced Grant OMNES (694544).

\bibliographystyle{SciPost_bibstyle}
\bibliography{QT}

\end{document}